\def \SAIT #1 #2 {{\em Mem.\ Soc.\ Astron.\ It.\/} {\bf #1}, #2}
\def \MESS #1 #2 {{\em The Messenger\/} {\bf #1}, #2}
\def \ASTRNACH #1 #2 {{\em Astron. Nach.\/} {\bf #1}, #2}
\def \AAP #1 #2 {{\em Astron. Astrophys.\/} {\bf #1}, #2}
\def \AAL #1 #2 {{\em Astron. Astrophys. Lett.\/} {\bf #1}, L#2}
\def \AAR #1 #2 {{\em Astron. Astrophys. Rev.\/} {\bf #1}, #2}
\def \AAS #1 #2 {{\em Astron. Astrophys. Suppl. Ser.\/} {\bf #1}, #2}
\def \AJ #1 #2 {{\em Astron. J.\/} {\bf #1}, #2}
\def \ANNREV #1 #2 {{\em Ann. Rev. Astron. Astrophys.\/} {\bf #1}, #2}
\def \APJ #1 #2 {{\em Astrophys. J.\/} {\bf #1}, #2}
\def \APJL #1 #2 {{\em Astrophys. J. Lett.\/} {\bf #1}, L#2}
\def \APJS #1 #2 {{\em Astrophys. J. Suppl.\/} {\bf #1}, #2}
\def \APSS #1 #2 {{\em Astrophys. Space Sci.\/} {\bf #1}, #2}
\def \ASR #1 #2 {{\em Adv. Space Res.\/} {\bf #1}, #2}
\def \BAIC #1 #2 {{\em Bull. Astron. Inst. Czechosl.\/} {\bf #1}, #2}
\def \JSQRT #1 #2 {{\em J. Quant. Spectrosc. Radiat. Transfer\/} {\bf #1}, #2}
\def \MN #1 #2 {{\em Mon. Not. R. Astr. Soc.\/} {\bf #1}, #2}
\def \MEM #1 #2 {{\em Mem. R. Astr. Soc.\/} {\bf #1}, #2}
\def \PLR #1 #2 {{\em Phys. Lett. Rev.\/} {\bf #1}, #2}
\def \PASJ #1 #2 {{\em Publ. Astron. Soc. Japan\/} {\bf #1}, #2}
\def \PASP #1 #2 {{\em Publ. Astr. Soc. Pacific\/} {\bf #1}, #2}
\def \NAT #1 #2 {{\em Nature\/} {\bf #1}, #2}

\def\spose#1{\hbox to 0pt{#1\hss}}
\def\simlt{\mathrel{\spose{\lower 3pt\hbox{$\mathchar"218$}}
     \raise 2.0pt\hbox{$\mathchar"13C$}}}
\def\simgt{\mathrel{\spose{\lower 3pt\hbox{$\mathchar"218$}}
     \raise 2.0pt\hbox{$\mathchar"13E$}}}
\def\eg{{\rm e.g.}}
\def\ie{{\rm i.e.}}
\def\etal{{\rm et~al.}}

\documentstyle{memsait}
\input epsf.sty
\begin{opening}
\title{THE DISTANCE TO THE LARGE MAGELLANIC CLOUD}
\author{BRAD K. GIBSON$^1$}
\institute{$^1$Center for Astrophysics \& Space Astronomy,
Univ. of Colorado, Boulder, CO  80309-0389}
\date{}
\end{opening}

\begin{document}

\oddpagefooter{}{}{}
\evenpagefooter{}{}{}
\ 
\bigskip

\begin{abstract}
A summary of recent Large Magellanic Cloud distance determinations $\mu_{\rm 
LMC}$
is presented, with an eye towards pinpointing the source(s) of the resulting
large discrepancies encountered between some of the techniques.  Thirty-eight
recent (1998-1999) measurements of $\mu_{\rm LMC}$
are highlighted, the extrema for
which (18.07 versus 18.74) are inconsistent with one another at the
$\simgt$3$\sigma$ level.  The lack of overlap between the results of many of the
techniques, simply employing the authors' quoted uncertainties, is a clear
indication that unrecognized uncertainties, both random and systematic,
plague many of the published results.  
While $\mu_{\rm LMC}$ 
almost certainly lies within $\sim$13\% of 18.5 (\ie, between
18.20 and 18.75), to those of us outside the LMC ``community'', no single
compelling argument has been put forth that reconciles 
the wildly disparate values presented thus far.  A $\sim$13\%
uncertainty in the LMC distance 
corresponds to a $\sim$13\% uncertainty in the Cepheid-based extragalactic
distance scale.
\end{abstract}

\vspace{-3mm}
\section{Introduction}
\vspace{-2mm}

A precise 
(and accurate!) value of the distance modulus 
to the LMC, $\mu_{\rm LMC}$, is a crucial 
component of the current Cepheid-based extragalactic distance scale;
$\mu_{\rm LMC}$ provides the anchor against which \it all \rm HST-based Cepheid
distance determinations are measured.  To date, both the \it HST Key Project on
the Extragalactic Distance Scale \rm (Gibson \etal\ 2000; Mould \etal\ 2000) 
and the \it
Sandage/Saha Type Ia Supernovae Calibration Team \rm (Saha \etal\ 1999) have
adopted the ``canonical'' zero point of $\mu_{\rm LMC}$=18.50 - the uncertainty
in $\mu_{\rm LMC}$ adopted by the \it Key Project \rm was $\pm 0.13$\,mag (\ie,
$\pm$6.5\%), representing the largest component of their global systematic
error budget.  While $\pm 0.13$\,mag may be the formal
uncertainty associated with the ensemble distribution of published $\mu_{\rm
LMC}$
values (in agreement with a similar analysis done by Jha \etal\ 1999), I
hesitate to attach a high level of confidence to such statistical analyses 
since the individual $\mu_{\rm LMC}$ ``components'' in the ensemble
distribution are as different as apples and oranges, with an uncomfortable lack
of internal consistency between the apples and apples (and oranges and
oranges)!

In what follows, I present a cursory overview of current (mid-Oct 1999)
thinking regarding the
distance to the LMC, contrasting the techniques employed in determining
$\mu_{\rm LMC}$, and commenting upon their respective strengths and weaknesses.
The implications for the extragalactic distance scale are inexorably linked and
noted in Section~3.
Complementary recent reviews include Popowski \& Gould (1999),
Layden (1999), and Feast (1999).

\vspace{-3mm}
\section{The Distance to the LMC}
\vspace{-2mm}

Figure~1 summarizes the state of the field at the time of writing 
(mid-Oct 1999), and should be considered an updated, and somewhat more 
comprehensive, companion to Cole's (1998) Figure~1.  Thirty-eight entries have
been included here, and are displayed in descending order of 
$\mu_{\rm LMC}$.  The technique
employed is listed to the right of each entry, along with a reference
number matching those in the caption.  

\begin{figure}
\epsfxsize=14cm
\hspace{-0.65cm}\epsfbox{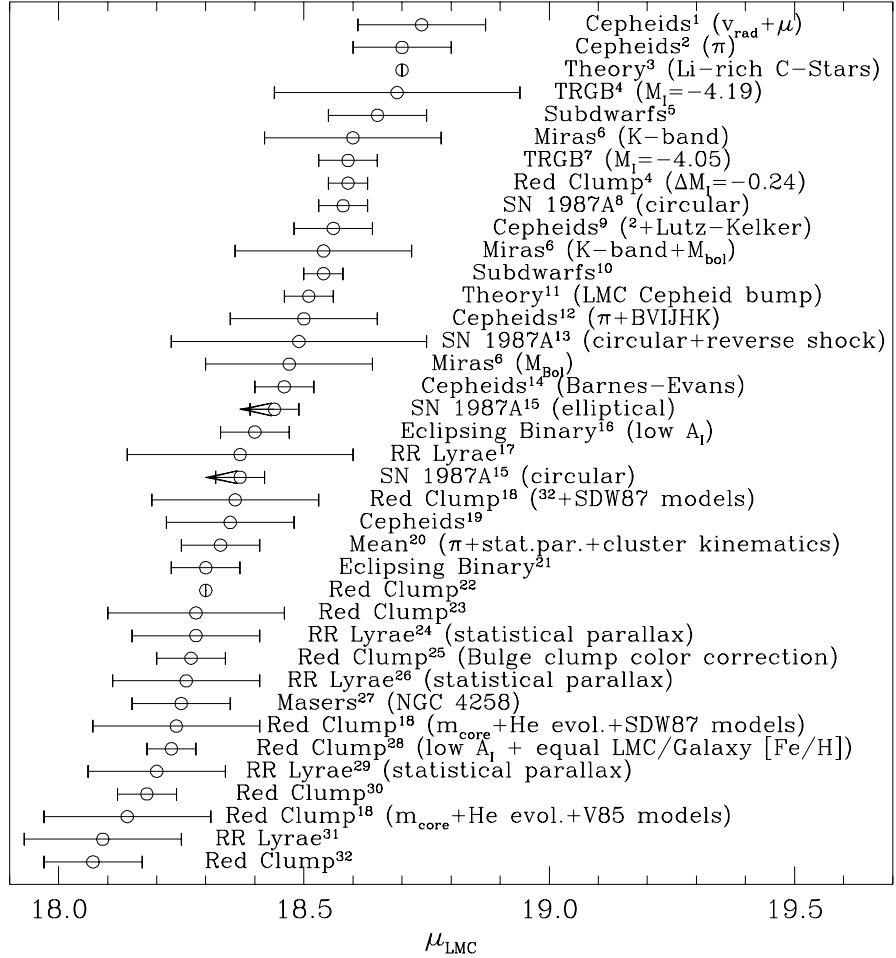}
\vspace{-0.7cm}
\caption[h]{Compilation of recent distance determinations to the LMC, presented
in decreasing order of modulus $\mu_{\rm LMC}$.  Cepheids, fitting to local
Galactic subdwarf sequences, and theoretical stellar models
tend to favor the ``long'' distance scale (\ie, $\mu_{\rm LMC}
\simgt 18.5$, while RR Lyrae, red clump luminosities, eclipsing binaries, and
masers (indirectly, through NGC~4258) tend to favor the
``short'' scale (\ie, $\mu_{\rm LMC} \simlt 18.4$).
References: 
$^{1}$Feast \etal\ (1998); 
$^{2}$Feast \& Catchpole (1997);
$^{3}$Ventura \etal\ (1999);
$^{4}$Romaniello \etal\ (1999);
$^{5}$Reid (1997);
$^{6}$van~Leeuwen \etal\ (1998);
$^{7}$Sakai \etal\ (2000);
$^{8}$Panagia (1998);
$^{9}$Oudmaijer \etal\ (1998);
$^{10}$Caretta \etal\ (1999);
$^{11}$Wood (1998);
$^{12}$Madore \& Freedman (1998);
$^{13}$Garnavich \etal\ (1999);
$^{14}$Gieren \etal\ (1998);
$^{15}$Gould \& Uza (1998);
$^{16}$Nelson \etal\ (1999);
$^{17}$Luri \etal\ (1998);
$^{18}$Cole (1998);
$^{19}$Luri \etal\ (1999);
$^{20}$Popowski \& Gould (1999);
$^{21}$Guinan \etal\ (1998);
$^{22}$Beaulieu \& Sackett (1998);
$^{23}$Girardi \etal\ (1998);
$^{24}$Layden \etal\ (1996);
$^{25}$Fernley \etal\ (1998);
$^{26}$Popowski (1999);
$^{27}$Maoz \etal\ (1999);
$^{28}$Udalski (1999);
$^{29}$Popowski \& Gould (1998) and Gould \& Popowski (1998);
$^{30}$Udalski (1998b);
$^{31}$Udalski (1998a);
$^{32}$Stanek \etal\ (1998).}
\end{figure}

Taking the extrema at face value ($\mu_{\rm LMC}$=18.07 and 18.74)
demonstrates the gravity of the situation for the extragalactic distance scale.
For example, assuming $\mu_{\rm LMC}$=18.50, Gibson \etal\ (2000) derived
H$_\circ$=68\,km\,s$^{-1}$\,Mpc$^{-1}$; for $\mu_{\rm LMC}$=18.07, 
though,\hfil\break
H$_\circ$=83\,km\,s$^{-1}$\,Mpc$^{-1}$, and for $\mu_{\rm LMC}$=18.74,
H$_\circ$=61\,km\,s$^{-1}$\,Mpc$^{-1}$ - \ie, H$_\circ$ could change by up to
$\sim$20\%, simply by revising the choice of $\mu_{\rm LMC}$!

A few general trends can be gleaned from a cursory inspection of Figure~1.
First, methods based upon Cepheids and fitting to local subdwarf sequences \it
generally \rm tend to favor the ``long'' distance scale (\ie,
$\mu_{\rm LMC}$$\simgt$18.5).  Second, those based upon RR~Lyrae statistical
parallax and luminosity of the LMC red clump stars \it generally \rm tend to
favor the ``short'' scale (\ie, $\mu_{\rm 
LMC}$$\simlt$18.4).  Third, those based
upon the exact same data (\eg, SN~1987A circumstellar ring geometry and the LMC
eclipsing binary HV~2274) are disturbingly author-dependent.

\vspace{-3mm}
\subsection{Cepheids}
\vspace{-2mm}

The most extreme proponents of the long distance scale are
Feast \etal\ (1998) and Feast \& Catchpole (1997), with
$\mu_{\rm LMC}$$\simgt$18.7.  Their earlier value of
$\mu_{\rm LMC}$=$18.70\pm 0.10$,
based upon HIPPARCOS trigonometric parallaxes of Galactic Cepheids, 
coupled with the $V$-band photometry of Caldwell \& Laney (1991), has
been re-examined by three different groups and suggested to be too large by 
$0.13-0.38$\,mag.  Madore \& Freedman (1998) show that full
multi-color photometry (adopting permutations of two to six photometric bands
for various Cepheid subsets) leads to a downward revision of $0.13-0.36$\,mag.
Oudmaijer \etal\ (1998) retain the single $V$-band
photometry that Feast \& Catchpole adopted, but apply a 
Lutz-Kelker bias correction to the parallaxes resulting in a 
0.14\,mag downward revision.  
Luri \etal\ (1998) employed their
``LM'' maximum-likelihood model (which incorporates proper motions $\mu$,
radial velocities $v_{\rm rad}$, and trigonometric parallaxes $\pi$, into the
mix, although $\pi$ enters in with apparently minimal weight -
Feast 1999) to revise Feast \& Catchpole downward by 
0.38\,mag; this represents the lowest Cepheid-based determination of
$\mu_{\rm LMC}$, $18.35\pm 0.13$.  In contrast with Luri \etal, 
Feast \etal\ (1998), also using a sample culled by $v_{\rm rad}$ and $\mu$,
find $\mu_{\rm 
LMC}$=18.74$\pm$0.13; the source of the discrepancy between the two
studies is not readily apparent.

\vspace{-3mm}
\subsection{Subdwarf Fitting}
\vspace{-2mm}

Subdwarf fitting techniques give the distance to a globular cluster by matching
the cluster sequence to a corresponding one derived from calibrating stars
in the solar neighborhood; local subdwarfs with HIPPARCOS parallaxes
provide the calibrating sample.  
Knowledge of the cluster's distance then yields an
absolute luminosity for the resident RR~Lyrae
(Popowski \& Gould 1999, equation~18).  Assuming a range of cluster
metallicities spanning that of the LMC RR~Lyrae can be sampled, one can derive
$\mu_{\rm LMC}$.  
As Figure~1 shows, subdwarf fitting techniques generally lead to
a $\mu_{\rm LMC}$ consistent with the long distance scale.
Reid (1997) and Caretta \etal\ (1999) claim $\mu_{\rm LMC}$$\approx
18.55-18.65$, a consequence of which is that
cluster horizontal branch stars are predicted to be
$\sim$0.2\,mag more luminous than their
field counterparts.  Catelan (1998) has questioned
the reality of this supposed cluster/field horizontal branch
luminosity difference.
Reid's analysis leads to an extremely steep slope
for the RR~Lyrae M$_{\rm V}$-[Fe/H] relationship; his
Figure~10 suggests a slope of $0.57\pm 0.35$, which seems difficult
to reconcile with that observed in Galactic and M31 globular cluster systems
(\ie, $0.18\pm 0.03$; Fusi-Pecci \etal\ 1996).  
Gould \& Popowski (1998a,b) and Popowski \& Gould (1999) have stressed the 
influence of
differing metallicity scales in the analyses as the primary source of the
discrepancy between their ``short'' (faint horizontal branch (M$_{\rm V}\approx
0.74$ at [Fe/H]=$-$1.6) and older ($t\simgt 16$\,Gyrs) globulars) RR~Lyrae
scale and the Reid/Caretta et~al. ``long'' (bright horizontal branch (M$_{\rm
V}\approx 0.47$ at [Fe/H]=$-$1.6) and younger ($t\approx 13$\,Gyrs) globulars)
scale.  Pinsonneault \etal\ (1998) suggest that residual spatially-dependent 
systematic HIPPARCOS parallax errors should also be considered as a potential
source of bias.

\vspace{-3mm}
\subsection{Statistical Parallax}
\vspace{-2mm}

Statistical parallax works on the principle that one can balance the stellar
radial velocity distribution of a sample of Galactic RR~Lyrae (distance
independent) against that of
its stellar proper motion distribution (distance dependent).  The balanced 
velocity ellipsoids provide the necessary distance scale parameter, which in
turns provides the necessary RR~Lyrae absolute luminosity calibration.
Statistical parallax, both prior to and subsequent to HIPPARCOS, has
consistently supported the ``short'' distance scale (\eg, Layden \etal\ 1996;
Gould \& Popowski 1998a,b; Layden 1999).

\vspace{-3mm}
\subsection{Miras}
\vspace{-2mm}

Distances based upon linking Miras in the LMC with those in the Galaxy 
with accurate HIPPARCOS parallaxes, because they are based on such a small
sample (11 overtone pulsators, culled from an initial total sample of
15 candidates - van Leeuwen \etal\ 1997), still have a large intrinsic
scatter ($\pm$0.18\,mag) and suffer from a 0.13\,mag systematic uncertainty
depending upon the bandpass adopted.
The ability to separate fundamental mode pulsators
from overtone pulsators is also crucial, as any contamination by the former
will systematically decrease the predicted $\mu_{\rm LMC}$.

\vspace{-3mm}
\subsection{SN~1987A}
\vspace{-2mm}

An appealing one-step
technique which bypasses the intermediate assumption that various Galactic 
stellar constituents (\eg, RR~Lyrae, Cepheids, Miras) are direct analogs
to those in the LMC, is that based
upon the geometry of the fluorescent light echo from the
SN~1987A circumstellar ring.  The technique consists
of measuring the angular size of the ring with HST and comparing it to
the absolute size, as estimated from the IUE emission line light
curves.  In terms of observables, the distance $d$ can be written (after
Panagia 1998):
\begin{equation}
d={{c(t_\circ+t_{\rm max})}\over{2\tan R^{\prime\prime}}} \quad[{\rm cm}],
\end{equation}
\noindent
where $t_\circ=86$\,d is the onset of UV emission, $t_{\rm max}$ is the
time of maximum UV line emission, and $R^{\prime\prime}$ is the apparent
angular ring size at $t_{\rm max}$.  It is in the assignment of values to
$t_{\rm max}$ and $R^{\prime\prime}$ where things start to get controversial.

Gould \& Uza (1998) adopt the weighted mean of the Plait \etal\ (1995) 
HST [OIII] ring angular size determinations (\ie, 
$R^{\prime\prime}=0.858\pm 0.011$\,arcsec) and the IUE NIII] and
NIV] light curves to derive $t_{\rm max}=378\pm 5$\,d.  
Panagia (1998) though attaches weight to the Jakobsen \etal\ (1991)
1988 pre-COSTAR 
[OIII] observation of 0.830$\pm$0.015, and uses
this in combination with the Plait \etal\ data to extrapolate back to
$t_{\rm max}$, yielding $R^{\prime\prime}=0.808\pm 0.017$.
Panagia claims the ring radius grew in extent by 6\% between 1988 and
1993, while Gould \& Uza assume a static ring.
Panagia uses the CIII], OIII], and NIII] IUE light curves
and favors $t_{\rm max}=395\pm 5$\,d.  
Both $t_{\rm max}$ solutions, despite being formally inconsistent at the
2$\sigma$ level, look equally good to the eye.  
Adopting their respective values, we can
use equation~1 to derive $\mu_{\rm LMC}$=18.37$\pm$0.05 (Gould \& Uza 1998) and
$\mu_{\rm LMC}$=18.55$\pm$0.05 (Panagia 1998 - aside: Panagia applies a $+$0.03\
shift to this value, under the assumption that SN~1987A lies 700\,pc in front of
the LMC's center of mass).  One-third of the resulting 0.18\,mag discrepancy 
is due to the different $t_{\rm max}$ adopted, with
the remaining two-thirds due to $R^{\prime\prime}$.
If the ring is intrinsically elliptical, as suggested by Crotts (as cited
by Gould \& Uza), 
both teams' $\mu_{\rm LMC}$ may be underestimated by up to 0.07\,mag.
The recent H$\alpha$+NII observation of the reverse shock by Garnavich
\etal\ (1999) has not (yet) clarified the above controversy.

Gould \& Uza claim that these $\mu_{\rm LMC}$ determinations are
only \it upper limits\rm, due to the inherent ``prompt
response assumption'' employed (\ie, that no time delay exists between the
blast of ionizing photons impinging upon the ring and the 
subsequent UV fluorescence).  This
fact has been used to reconcile the inferred SN~1987A-based $\mu_{\rm LMC}$ 
with 
the lower value derived via the LMC's red clump and masers in NGC~4258 
(discussed below), which both yield $\sim$18.25.
What is not appreciated in these arguments is the 
magnitude of the putative time delay needed.  Specifically, from equation~1,
$\Delta\mu_{\rm LMC}$$\approx -0.005$\,mag for every day of ``delay'', 
independent of $R^{\prime\prime}$.  

In practice, what this means is that to reconcile Gould \& Uza's
results (assuming an intrinsically circular ring) with $\mu_{\rm 
LMC}$=18.25, one
would need to reduce $t_{\rm max}$ from 378 to 357\,d (\ie, a time delay of
21\,d).  The situation is even worse when attempting to reconcile Panagia
with $\mu_{\rm LMC}$=18.25; in that case, $t_{\rm max}$ would have to be
reduced from 395 to 331\,d (\ie, a time delay of 64\,d).  Are delays of
this enormous magnitude physically feasible?  I have not seen this particular
scenario addressed, but an informal polling of my colleagues with experience in
such fluorescence
models suggests not - ``a few days'' was the most optimistic
response, meaning the downward correction to $\mu_{\rm LMC}$
could be no more than
$\sim$0.01$-$0.02\,mag.  Relaxing the prompt-response assumption, therefore,
does not appear to be the panacea it might at first appear to be.
Continuing to refer to the SN~1987A results as ``upper limits'' is somewhat
misleading, in my opinion.

\vspace{-3mm}
\subsection{Red Clump}
\vspace{-2mm}

Under the assumption that the mean $I$-band luminosity of red clump stars in the
LMC matches those calibrated locally by HIPPARCOS, Udalski (1998b) and
Stanek \etal\ (1998) found $\mu_{\rm LMC}$$\approx$18.1.  Cole (1998) criticized
these analyses on the basis that age and metallicity differences between the
LMC and Galactic red clump stars cannot be neglected.  By employing Seidel
\etal\ (1987) red clump models, evolved under the (poor) assumption of constant
helium abundance and core mass, Cole claimed that Udalski and Stanek \etal\
underestimated $\mu_{\rm LMC}$ by 0.28\,mag.  This
extreme correction though is not supported by modern stellar
evolution models (\eg, Girardi \etal\ 1998).  In fact, even in comparison with
its contemporaries (\eg, VandenBerg 1985; Lattanzio 1986; Sweigert \& Gross
1978), the Seidel \etal\ models are extreme.  Adoption of any of 
the other models of that era would have led to a predicted correction factor of
0.1$\pm$0.1, as opposed to the $\sim$0.3 suggested by Cole.
Empirically,
Stanek \& Garnavich (1998) show that the stellar populations in
three M31 fields have identical $I$-band luminosities.  Considering the
galactocentric distances of the fields (6.7, 11.2, and 33.6\,kpc) with 
the Zaritsky \etal\
(1994) abundance gradient ($-$0.018\,dex/kpc), one would predict an \it
a priori \rm metallicity differential of $\sim$0.5\,dex, in agreement with 
each field's CMD (Holland \etal\ 1996; Rich
\etal\ 1996).  In the parlance of Cole, this
0.5\,dex difference implies that there should be a 
0.15$\pm$0.05\,mag
difference in the $I$-band luminosities of the inner and outer M31
field CMDs - such
a difference is not observed.
Further evidence to the minimal role played by age differences
is provided in Udalski (1998b), who showed that the $I$-band luminosities of
LMC and SMC clusters was independent of age (for cluster ages of
$\sim$2$-$10\,Gyrs).

More recently, both Zaritsky (1999) and Romaniello \etal\ (1999) have suggested
an upward revision of $\sim$0.2\,mag be applied to 
all previous red clump determinations, based upon their
more sophisticated treatment of dust in the LMC red clump fields.  In response,
Udalski (1999) bypasses the complicated dust corrections favored by
Zaritsky and Romaniello \etal, by simply restricting the analysis to regions of
low extinction, thereby obviating any complicated dust corrections.  Udalski
also makes the point that the region in which Romaniello \etal\ restricted
their red clump analysis is unsuitable for red clump studies.  Of perhaps even
greater importance, Udalski culled from the HIPPARCOS Galactic
red clump catalog, a large sample of stars \it with LMC metallicities \rm (from
McWilliam 1990), thereby avoiding any potential 
Galactic versus LMC stellar population mismatch pitfalls.  His result -
$\mu_{\rm LMC}$=18.23$\pm$0.05 - appears to be the most solid empirical
application of this technique possible, at this time.

\vspace{-3mm}
\subsection{Tip of the Red Giant Branch (TRGB)}
\vspace{-2mm}

Romaniello \etal\ (1999) and Sakai \etal\ (2000) have derived 
$\mu_{\rm 
LMC}$$\approx$18.6$-$18.7, based upon the apparent magnitude of
the TRGB, coupled with the assumption that the $I$-band luminosity of the 
TRGB (for the LMC's metallicity) 
corresponds to M$_I\approx -4.1\rightarrow -4.2$.  Udalski (1999) has already
commented upon the suitability and photometric calibration 
of the field employed by Romaniello \etal.  To reconcile the Sakai \etal\ value
of $\mu_{\rm LMC}$=18.59
with $\mu_{\rm LMC}$=18.25 would require M$_I\approx -3.7$.

\vspace{-3mm}
\subsection{Eclipsing Binaries}
\vspace{-2mm}

The geometry of the LMC eclipsing binary system HV~2274 has been used by Guinan
\etal\ (1998) to derive $\mu_{\rm LMC}$=18.30$\pm$0.07.  This 
technique, while not entirely geometrical (local reddening estimates are
required), remains one of the most promising.  A
graphic illustration of the importance of using the appropriate reddening comes
from the fact that Guinan
\etal\ adopted a reddening of E(B$-$V)=0.12$\pm$0.01, while
in contrast, Nelson \etal\ (1999) derived a reddening of
E(B$-$V)=0.09$\pm$0.03, leading to $\mu_{\rm LMC}$=18.40$\pm$0.07.

\vspace{-3mm}
\subsection{Masers in NGC~4258}
\vspace{-2mm}

Under the assumption that $\mu_{\rm LMC}$=18.50$\pm$0.13, 
Maoz \etal\ (1999) derived a Cepheid
distance to NGC~4258 of 29.54$\pm$0.12$_{\rm r}$$\pm$0.18$_{\rm s}$, where the
subscripts 'r' and 's' correspond to random and systematic uncertainties,
respectively.  Conversely, Herrnstein \etal\ (1999) derived a purely geometric
distance to NGC~4258 of 29.29$\pm$0.10$_{\rm r}$$\pm$0.12$_{\rm s}$,
using the orbital motions of masers in the gas disk
surrounding the galaxy's nucleus.  Unless some unappreciated flaw exists in 
either (or both) of these analyses, in order to reconcile the
Cepheid and maser distances, one must reduce $\mu_{\rm LMC}$ from 18.50 to
$\sim$18.25.

\vspace{-3mm}
\section{Summary}
\vspace{-2mm}

To re-iterate points made earlier, it is clear from even a cursory examination
of Figure~1 that significant unappreciated systematic 
uncertainties exist in many of the published values of $\mu_{\rm LMC}$, 
as evident
in the lack of overlap in the quoted uncertainties.  I am loathe to pass
judgment on the various entries shown in Figure~1, but will
mention that those techniques deemed as ``geometric'' (and, in principle,
somewhat less susceptible to intermediate steps in going from Galactic to LMC
stellar populations), such as the eclipsing binary HV~2274, the circumstellar
ring around SN~1987A (at least Gould \& Uza's analysis), 
and the masers in NGC~4258, all suggest $\mu_{\rm 
LMC}$$\approx$18.3.  Cepheids (except for Luri \etal\ 1999) and subdwarf 
sequences still favor the long scale ($\mu_{\rm 
LMC}$$\approx$18.5$-$18.7).  I remain skeptical
that our understanding of $\mu_{\rm LMC}$ 
is accurate to 6\%, as has been assumed in
most Cepheid-based extragalactic distance scale analyses, and would
suggest that 13\% is a fairer reflection of the present-day uncertainty.
I stress that this 13\% should \it not \rm be construed in terms of any sort of
Gaussian confidence level (\eg, 1$\sigma$, etc.); I do not currently support
the use of $\mu_{\rm LMC}$ distributions as useful statistical measures of the
uncertainty in $\mu_{\rm LMC}$.  

Future astrometric missions such as FAME ({\tt
http://aa.usno.navy.mil/FAME/\rm}) and SIM ({\tt http://sim.jpl.nasa.gov/\rm})
should provide a definitive calibration for
the absolute luminosities of Galactic Cepheid and RR~Lyrae standard candles,
and hopefully resolve the remaining discrepancies, one way or the other.

\vspace{-3mm}
\acknowledgements
\vspace{-2mm}
I wish to thank both Francesca D'Antona and Roberto Gallino for their
assistance in ensuring that this contribution was included in these
proceedings.  Helpful feedback from Mike Shull, Dick McCray, 
Bohdan Paczynski, and
Andrzej Udalski was likewise appreciated.  BKG acknowledges support from the
NASA Long-Term Space
Astrophysics Program (NAG5-7262) and the FUSE Science Team (NAS5-32985).

I want to dedicate this paper to the memory of my good friend Charlene Heisler
- you are sorely missed Charlene.

\end{document}